\documentclass[prd,aps, tightenlines, preprintnumbers, showpacs, nofootinbib,superscriptaddress,notitlepage]{revtex4-1}

\usepackage{graphicx}
\usepackage{amsmath,amsthm,amssymb}

\def\empile#1\above#2{\mathrel{\mathop{\kern 0pt#1}\limits_{#2}}}

\newcommand{\sll}{\raise.15ex\hbox{$/$}\kern-.43em\hbox{$l$}}
\newcommand{\slepsilon}{\raise.15ex\hbox{$/$}\kern-.53em\hbox{$\epsilon$}}
\newcommand{\slvarepsilon}{\raise.15ex\hbox{$/$}\kern-.53em\hbox{$\varepsilon$}}
\newcommand{\slL}{\raise.15ex\hbox{$/$}\kern-.53em\hbox{$L$}}
\newcommand{\slP}{\raise.15ex\hbox{$/$}\kern-.53em\hbox{$P$}}
\newcommand{\slp}{\raise.1ex\hbox{$/$}\kern-.63em\hbox{$p$}}
\newcommand{\slq}{\raise.1ex\hbox{$/$}\kern-.53em\hbox{$q$}}
\newcommand{\slv}{\raise.1ex\hbox{$/$}\kern-.63em\hbox{$v$}}
\newcommand{\slR}{\raise.15ex\hbox{$/$}\kern-.53em\hbox{$R$}}
\newcommand{\slQ}{\raise.15ex\hbox{$/$}\kern-.53em\hbox{$Q$}}
\newcommand{\slK}{\raise.15ex\hbox{$/$}\kern-.53em\hbox{$K$}}
\newcommand{\slk}{\raise.15ex\hbox{$/$}\kern-.53em\hbox{$k$}}
\newcommand{\slSigma}{\raise.15ex\hbox{$/$}\kern-.53em\hbox{$\Sigma$}}
\newcommand{\slcalP}{\raise.15ex\hbox{$/$}\kern-.63em\hbox{$\cal P$}}
\newcommand{\slA}{\raise.15ex\hbox{$/$}\kern-.73em\hbox{$A$}}
\newcommand{\slbfA}{\raise.15ex\hbox{$/$}\kern-.73em\hbox{${\imb A}$}}
\newcommand{\slpartial}{\raise.15ex\hbox{$/$}\kern-.53em\hbox{$\partial$}}
\newcommand{\sla}{\raise.15ex\hbox{$/$}\kern-.53em\hbox{$a$}}
\newcommand{\slb}{\raise.15ex\hbox{$/$}\kern-.53em\hbox{$b$}}
\newcommand{\slc}{\raise.15ex\hbox{$/$}\kern-.53em\hbox{$c$}}
\newcommand{\slD}{\raise.15ex\hbox{$/$}\kern-.53em\hbox{$D$}}
\newcommand{\slC}{\raise.15ex\hbox{$/$}\kern-.53em\hbox{$C$}}

\def\wt{\widetilde}

\begin{document}

\title{
Forward Heavy Quarkonium Productions at the LHC
}

\author{Kazuhiro Watanabe}
\affiliation{Key Laboratory of Quark and Lepton Physics (MOE) and Institute
of Particle Physics, Central China Normal University, Wuhan 430079, China}

\author{Bo-Wen Xiao}
\affiliation{Key Laboratory of Quark and Lepton Physics (MOE) and Institute
of Particle Physics, Central China Normal University, Wuhan 430079, China}

\begin{abstract}

We investigate the low transverse momentum heavy quarkonium ($J/\psi$, $\Upsilon$) productions in the 
forward rapidity region of $pp$ and $pA$ collisions at the LHC as an important probe to the transverse 
momentum tomography of the gluons in hadrons in the Color Glass Condensate (CGC) framework. 
By implementing the Sudakov resummation consistently in the CGC formalism, we 
achieve an excellent agreement between the improved CGC calculations and the LHC data. 
We show that both the small-$x$ and the Sudakov effects are essential for a complete description of 
heavy quarkonium productions in the low transverse momentum region. This
provides a solid foundation to study the small-$x$ gluon saturation in a big nucleus with
the future $pA$ programs at the LHC. 

\end{abstract}

\maketitle

\textit{Introduction}--
Due to the rich and complex dynamics of strong interaction and its close connection to various interesting 
phenomena, heavy quarkonium production in high energy collisions has been one of the most intriguing subjects in 
the study of strong interaction physics. In particular, as the objective of this paper, it has been shown that the 
low-$p_\perp$ spectrum of forward rapidity heavy quarkoniums produced in high 
energy $pp$ ($pA$) collisions carries important information of the gluon saturation 
at small-$x$. The concept of gluon saturation~\cite{Gribov:1984tu, Mueller:1985wy} 
comes from the anticipation of gluon recombination when its density becomes 
extremely high in high energy collisions. The associated dynamics is one
of the challenging physics topics in current and future nuclear science
program around the world~\cite{Accardi:2012qut}.

There have been intensive theoretical studies on heavy quarkonium productions in the small-$x$ 
saturation framework, i.e., the Color-Glass-Condensate (CGC) formalism. They have shown
that the typical transverse momentum of produced quarkonium is  around the saturation momentum $Q_s(x_g)$~\cite{Blaizot:2004wv,Kovchegov:2006qn,Fujii:2006ab,Kharzeev:2008cv, Fujii:2013gxa, Kang:2013hta, Ma:2014mri, Ma:2015sia, Ducloue:2015gfa} . As a function of the longitudinal momentum fraction of target hadron $x_g$, 
the saturation momentum increases when $x_g$ decreases, and it characterizes the typical transverse momentum that small-$x$ gluons carry in a hadron. It can also be derived from the boundary which separates the dilute regime from the dense saturated regime where gluons evolve non-linearly. The non-linear evolution equation at small-$x$ (or high energy) is known as the Balitsky-Kovchegov (BK) equation~\cite{Balitsky:1995ub, Kovchegov:1996ty}, which has been implemented numerically in CGC formalism. 

The physical picture of quarkonium productions in the CGC framework is as follows. 
Either before or after interacting with the 
dense gluons in the target hadron, a large $x$ gluon from projectile proton fluctuates into a pair of heavy quarks, which eventually forms a bound state of quarkonium with transverse momentum $p_\perp$. In the low $p_\perp$ region, the main contribution of the transverse momentum broadening is due to 
gluon saturation effects. At forward rapidity $y$, the kinematics indicates 
that $x_g$ from the target side is much smaller than the one in the projectile. 
This is the ideal region for the study of small-$x$ physics. Therefore, the heavy quarkonium productions in the forward $pp$ and $pA$ collisions 
can provide an important probe for the gluon saturation
at small-$x$, whose tomography imaging has attracted strong interests in 
hadron physics community~\cite{Accardi:2012qut}. 

Thanks to the abundant data~\cite{Khachatryan:2010yr, Aaij:2011jh, Aad:2011sp, Chatrchyan:2011kc, 
LHCb:2012aa,Aaij:2013yaa,Abelev:2014qha, Abelev:2013yxa, Aaij:2013zxa} and wide kinematical 
coverage at the LHC, we are able to test the CGC/saturation physics calculations for $pp$ and $pA$ 
collisions by comparing them to experimental data. Early studies of $J/\Psi$ 
productions~\cite{Fujii:2013gxa, Ma:2014mri, Ma:2015sia, Ducloue:2015gfa} have shown that the 
CGC can provide decent description of the data, especially the transverse 
momentum distribution of the produced $J/\Psi$. Due to similar kinematics and physical parameters 
except for differences in mass, one expect that the $p_\perp$ broadening of $\Upsilon$ should be 
similar or slightly smaller than the one observed for  $J/\Psi$ due to saturation effects~\cite{Ducloue:2015gfa}. 
However, the LHCb experiments~\cite{Aaij:2011jh, LHCb:2012aa, Aaij:2013yaa} have 
found consistently larger mean $p_\perp$ for $\Upsilon$ for a wide range of rapidity windows,
as compared to that of $J/\psi$, as well as the theoretical calculations of~\cite{Ducloue:2015gfa}. 
In this paper, we will solve this puzzle.

We will show that one needs to take into account the 
Sudakov resummation together with the small-$x$ effect, in order to consistently 
describe the $J/\psi$ and $\Upsilon$ data. In a recent paper~\cite{Mueller:2012uf}, 
it has been demonstrated that the Sudakov resummation can be performed simultaneously with the 
small-$x$ resummation. 
Furthermore, Refs.~\cite{Mueller:2013wwa, Qiu:2013qka} computed the corresponding 
Sudakov double logarithms (in terms of $\alpha_s C_A \ln^2 ({M^2}/{p_\perp^2})$ with $M$ the 
quarkonium mass) for heavy quarkonium productions in hadron-hadron collisions in the 
collinear/CGC hybrid formalism. The results are consistent with those obtained in the 
Collins-Soper-Sterman (CSS) formalism~\cite{Collins:1984kg}. Based on the commonly 
used CGC calculations for heavy quarkoniums, we implement the Sudakov effects and 
compute the $p_\perp$ spectrum for both $J/\psi$ and $\Upsilon$. Since the typical 
transverse momentum of the produced quarkonium, namely the saturation momentum 
$Q_s\sim 1$-$2 \, \textrm{GeV}$, is not much smaller than the $J/\psi$ 
mass, the Sudakov effect is found to be negligible for $J/\psi$ productions at the LHC. On the other
hand, the Sudakov logarithms are important for $\Upsilon$ which consists of bottom quark with the mass 
around $4 \,\textrm{GeV}$. By performing the numerical calculation with the Sudakov factors, 
which provide extra $p_\perp$ broadening for $\Upsilon$ spectra, we find a complete  
agreement with the LHC data for both $J/\psi$ and $\Upsilon$ in the low-$p_\perp$ region. 
This shall provide an important step forward to study the small-$x$ gluon saturation  
in the future $pA$ program at the LHC.

\textit{The implementation of the Sudakov resummation in the CGC framework}--
In the forward rapidity region, transverse momentum of gluon coming from incident proton should be of order less than ${\cal O}(\Lambda_{\rm QCD})$,
which allow us to apply the usual collinear gluon distribution for the dilute incident proton side.
Therefore we can start with the heavy quark pair production cross section formula in $pp$ ($pA$) collisions for minimum bias events at leading order~\cite{Fujii:2013gxa} as follows
\begin{align}
\frac{d\sigma_{q\bar{q}}}{d^2q_{q\perp} d^2q_{\bar{q}\perp} dy_{q} dy_{\bar q}}
=
\frac{\alpha_s^2}{16\pi^2 C_F}
\int d^2k_\perp
\frac{\Xi_{\rm coll}(k_{2\perp},k_{\perp})}{k_{2\perp}^2}
\;
 x_1 G(x_1,\mu)
\; 
\phi_{{A},x_2}^{q\bar{q},g}(k_{2\perp},k_\perp),
\label{eq:cross-section-LN-coll}
\end{align}
where $q_{q\perp}$ ($q_{\bar{q}\perp}$) and $y_{q}$ ($y_{\bar q}$) are transverse momentum and rapidity for 
produced quark (antiquark) respectively.
$x_1$ ($x_2$) is the longitudinal momentum fraction of the dilute proton (the dense nucleus) carried by the incoming gluon.
For quarkonium production, we find $x_{1,2}={\sqrt{M^2+p_\perp^2}}/{\sqrt{s}}e^{\pm Y}$ within $2\rightarrow1$ kinematics
where $M$ is invariant mass of a quark and antiquark pair and $p_\perp=q_{q\perp}+q_{\bar{q}\perp}$ is transverse momentum of the pair.
$p_\perp$ is nothing but quarkonium transverse momentum.
$Y={1}/{2}\ln(({q_{q}^++q_{\bar{q}}^+})/({q_{q}^-+q_{\bar{q}}^-}))$ is the pair rapidity.
$x_1 G$ is the usual collinear gluon distribution function with $\mu$ being the factorization scale.
In the above formula, we have $k_{2\perp}=p_\perp$ due to momentum conservation.
The hard matrix element is given by
$\Xi_{\rm coll}
=
\Xi_{\rm coll}^{q\bar{q},q\bar{q}}
+
\Xi_{\rm coll}^{q\bar{q},g}
+
\Xi_{\rm coll}^{g,g}\;$
with
\begin{align}
\Xi_{\rm coll}^{q\bar{q},q\bar{q}}
&=
\frac{8z(1-z)}{(a_\perp^2+m^2)^2}
\left[
m^2+\left\{z^2+(1-z)^2\right\}a_\perp^2
\right]\; ,
\notag\\
\Xi_{\rm coll}^{q\bar{q},g}
&=
-\frac{16}{M^2(a_\perp^2+m^2)}
\left[
m^2+\left\{z^2+(1-z)^2\right\}a_\perp\cdot\left\{(1-z)q_{q\perp}-zq_{\bar{q}\perp}\right\}
\right]\; ,
\notag\\
\Xi_{\rm coll}^{g,g}
&=
\frac{8}{M^4}
\left[
M^2-2\left\{(1-z)q_{q\perp}-zq_{\bar{q}\perp}\right\}^2
\right]
\end{align}
where $a_\perp\equiv q_{q\perp}-k_\perp$.
$\Xi_{\rm coll}^{q\bar{q},q\bar{q}}$ represents the scattering of heavy quark pair produced from gluon splitting off the dense nucleus and $\Xi_{\rm coll}^{g,g}$ corresponds to the scattering of gluon before heavy quark pair splitting off the dense nucleus, whereas $\Xi_{\rm coll}^{q\bar{q},g}$ is derived from the interference between the above two cases.
In the above expressions, we have introduced the momentum fraction $z$ for the quark
$
z=
{q_q^+}/({q_q^++q_{\bar{q}}^+})
$
and $z_{\bar{q}}=1-z$ for the antiquark. The multipoint point function $\phi_{{A},{y_g}}^{q\bar{q},g}$ is constructed with the color singlet dipole amplitude as follows;
\begin{align}
\phi_{{A},{Y_g}}^{q\bar{q},g}(k_{2\perp},k_\perp)
&=
\overline{S}_\perp \; 
\int \frac{d^2x_\perp d^2y_\perp}{(2\pi)^4} e^{-ik_\perp\cdot x_\perp} e^{i(k_{2\perp}-k_\perp)\cdot y_\perp}
S_{{Y_g}}(x_\perp) \;
S_{{Y_g}}(y_\perp)\notag\\
&=
\overline{S}_\perp \; 
F_{{Y_g}}(k_\perp) \;
F_{{Y_g}}(k_{2\perp}-k_\perp),
\label{eq:3ptfn_2}
\end{align}
where we denote $\overline{S}_\perp=S_\perp\frac{N_c k_{2\perp}^2}{4 \alpha_s}$ with $S_\perp=\pi R_{\rm A}^2$ as the transverse area of target nucleus and $F_{{Y_g}}(k_\perp)\equiv\int \frac{dx_\perp^2}{(2\pi)^2} e^{-ik_\perp\cdot x_\perp} {S}_{{Y_g}}( x_\perp)$.
Here ${Y_g}=\ln\frac{x_0}{x_2}$ is the evolution rapidity of the gluon in the target with $x_0=0.01$.
The rapidity evolution of $S_{{Y_g}}$ is described by the nonlinear BK equation. For numerical calculations, we use the BK equation with running coupling correction (rcBK)~\cite{Balitsky:2006wa}.
The initial condition of the rcBK is the so-called McLerran-Venugopalan model~\cite{McLerran:1993ni} with modified anomalous dimension 
at $x_0=0.01$: 
\begin{align}
S_{{{Y_g}=0}}(x_\perp)=\exp\left[-\frac{(x_\perp^2Q_{s,0}^2)^\gamma}{4}\ln\left(\frac{1}{|x_\perp|\Lambda}+e\right)\right].
\label{eq;initial-condition}
\end{align}
It is well known that HERA e+p global data fitting~\cite{Albacete:2010sy,Fujii:2011fh,Albacete:2012xq} provides a nice constrained initial condition by using Eq.~(\ref{eq;initial-condition}). In this paper, since the results are not sensitive to the initial condition, we adopt the same parameters set for Eq.~(\ref{eq;initial-condition}) as found in Ref.~\cite{Fujii:2013gxa}.
In $pA$ collisions, we change the initial saturation scale as $Q_{sA,0}^2=cA^{1/3}Q_{s,0}^2$ with $c=0.5$~\cite{Dusling:2009ni}.

The differential cross section of the pair production with the Sudakov factor in momentum space is given by 
\begin{align}
\frac{d\sigma_{q\bar{q}}}{d^2q_{q\perp} d^2q_{\bar{q}\perp} dy_q dy_{\bar{q}}}
=
\frac{\alpha_s^2}{16\pi^2 C_F}
\int d^2l_\perp d^2k_\perp
\frac{\Xi_{\rm coll}(k_{2\perp},k_{\perp}-zl_\perp)}{k_{2\perp}^2}
\phi_{x_1,x_2}(k_{2\perp},k_\perp,l_\perp)
\label{eq:xsection_full}
\end{align}
with
\begin{align}
\phi_{x_1,x_2}(k_{2\perp},k_\perp,l_\perp)
=
\overline{S}_\perp 
F_{Y_g}\left(k_\perp\right)
F_{Y_g}\left(k_{2\perp}-k_\perp+l_\perp\right)
F_{\rm Sud}(l_\perp)
\end{align}
where we denote the Fourier transform of the Sudakov factor with the gluon pdf as follows
\begin{align}
F_{\rm Sud}(M,l_\perp)=\int \frac{d^2b_\perp}{(2\pi)^2}e^{-ib_\perp\cdot l_\perp} e^{-S_{\rm Sud}(M,b_\perp)}x_1G\left(x_1,\frac{c_0}{b_\perp}\right),
\label{eq:3ptfn_sud}
\end{align}
where $c_0=2e^{-\gamma_{\rm E}}$ with $\gamma_{\rm E}$ the Euler-constant and the factorization scale is chosen as $\mu=\frac{c_0}{v_\perp}$.

In Eq.~(\ref{eq:3ptfn_sud}), 
the exponent of the Sudakov factor is given by
\begin{align}
S_{\rm Sud}(M,b)=S_{\rm perp}(M,b_\star)+S_{\rm NP}(M,b)
\end{align}
where we have introduced $b_\star=b/\sqrt{1+(b/b_{\rm max})^2}$ by following the CSS formalism~\cite{Collins:1984kg} 
to separate the perturbative part ($b\ll b_{\rm max}$) from the nonpertubative part ($b>b_{\rm max}$).
$b_{\rm max}$ is a cutoff scale to separate these two regions. $S_{\rm perp}(Q,b)$ is calculable perturbatively at small $b$
which can be cast into 
\begin{align}
S_{\rm perp}(M,b)=\int_{c_0/b^2}^{M^2}\frac{d\mu^2}{\mu^2}\left[A\ln\left(\frac{M^2}{\mu^2}\right)+B\right]
\label{eq;pertbative_sudakov}
\end{align}
where 
the coefficient functions $A$ and $B$ have been calculated perturbatively 
$A=\sum\limits_{i=1} A^{(i)}\left(\frac{\alpha_s}{\pi}\right)^i$ and 
$B=\sum\limits_{i=1} B^{(i)}\left(\frac{\alpha_s}{\pi}\right)^i$ respectively.
For the one loop correction, they are given by $A^{(1)}=C_A$ and $B^{(1)}=-(b_0+\frac{1}{2}\delta_{8c})N_c$
where $b_0=\left(\frac{11}{6}N_c-\frac{n_f}{3}\right)\frac{1}{N_c}$.
As for $B^{(1)}$, the factor $\delta_{8c}$ is significant only in the production of a color octet quark-antiquark pair. In our calculation, because 
the quarkonium is produced by the color octet channel in the color evaporation model (CEM) with large-$N_c$,
we will take into account the contribution from this term.

For the non-perturbative part $S_{\rm NP}(\mu,b)$, we adopt the one from Ref.~\cite{Sun:2012vc} derived for the color octet quarkonium production as follows
\begin{align}
S_{\rm NP}(M,b)=\exp\left[\frac{b^2}{2}\left(-g_1-g_2\ln\left(\frac{M}{2Q_0}\right)-g_1g_3\ln(100x_1x_2)\right)\right]
\label{eq:npsudakov}
\end{align}
where $g_1=0.03$, $g_2=0.87$, and $g_1\times g_3=-0.17$ are obtained by the data fitting within the NRQCD factorization
with $Q_0=1.6~{\rm GeV}$ and $b_{\rm max}=0.5~{\rm GeV}$ is originally fixed in Ref.~\cite{Sun:2012vc}. Since we only employ the collinear gluon distribution for the large $x$ gluon from the incident proton, 
Eq.~(\ref{eq:npsudakov}) differs by a half as compared to Ref.~\cite{Sun:2012vc}. As found in Ref.~\cite{Sun:2012vc}, all the parameters are determined by assuming that the color octet heavy quark pair is dominant channel of
the quarkonium production at low transverse momentum. In the same spirit, we can use the same parameters for the CEM.


\textit{Numerical results}--
Now we turn to the numerical calculations for quarkonium production in the forward $pp$ and 
$pA$ collisions by adopting the CEM. We expect that the Sudakov effect should also apply in other models. In the CEM, the invariant mass of heavy quark pair is integrated from the bare quark pair mass to the threshold of decay into an open heavy flavor meson pair, namely $2m\leq M \leq 2M_{h}$.
We fix $m=1.2$ GeV and $M_{{h}=D}=1.864$ GeV for $J/\psi$ production, and $m=4.5$ GeV and $M_{{h}=B}=5.280$ GeV for $\Upsilon$ respectively.
The produced quark pair is going to be bound into a quarkonium with the probability $F_{q\bar{q}\rightarrow {\cal Q}}$.  This empirical factor is interpreted as a normalization factor because it depends on quark mass and factorization scale and etc. We have also included any $K$ factor in association with higher order correction in $F_{q\bar{q}\rightarrow {\cal Q}}$ as well.

Some setup parameters in numerical calculations are listed here.
We fix $R_{\rm p}=0.9~\rm{fm}$ and $R_{\rm A}=8.5~\rm{fm}$, $\alpha_s=0.3$ in the hard part. 
As for the running coupling $\alpha_s(Q)$ in $S_{\rm perp}(M,b_\star)$, 
we use $\Lambda_{\rm QCD}^{(4)}=0.173$ GeV for 4-flavor in $\Upsilon$ productions 
and $\Lambda_{\rm QCD}^{(3)}=0.214$ GeV for 3-flavor in $J/\psi$ productions.
These values are extracted to reproduce the low-$Q$ data of $\alpha_s(Q)$~\cite{Agashe:2014kda} 
up to $Q=10$ GeV by using 1-loop running coupling.
In addition, we adopt CTEQ6 pdf~\cite{Pumplin:2002vw} in numerical calculations:
CTEQ6L for Eq.~(\ref{eq:cross-section-LN-coll})
and CTEQ6M for Eq.~(\ref{eq:xsection_full}).

\begin{figure}[tbp]
 \centering
 \includegraphics[height=5.9cm,angle=270]{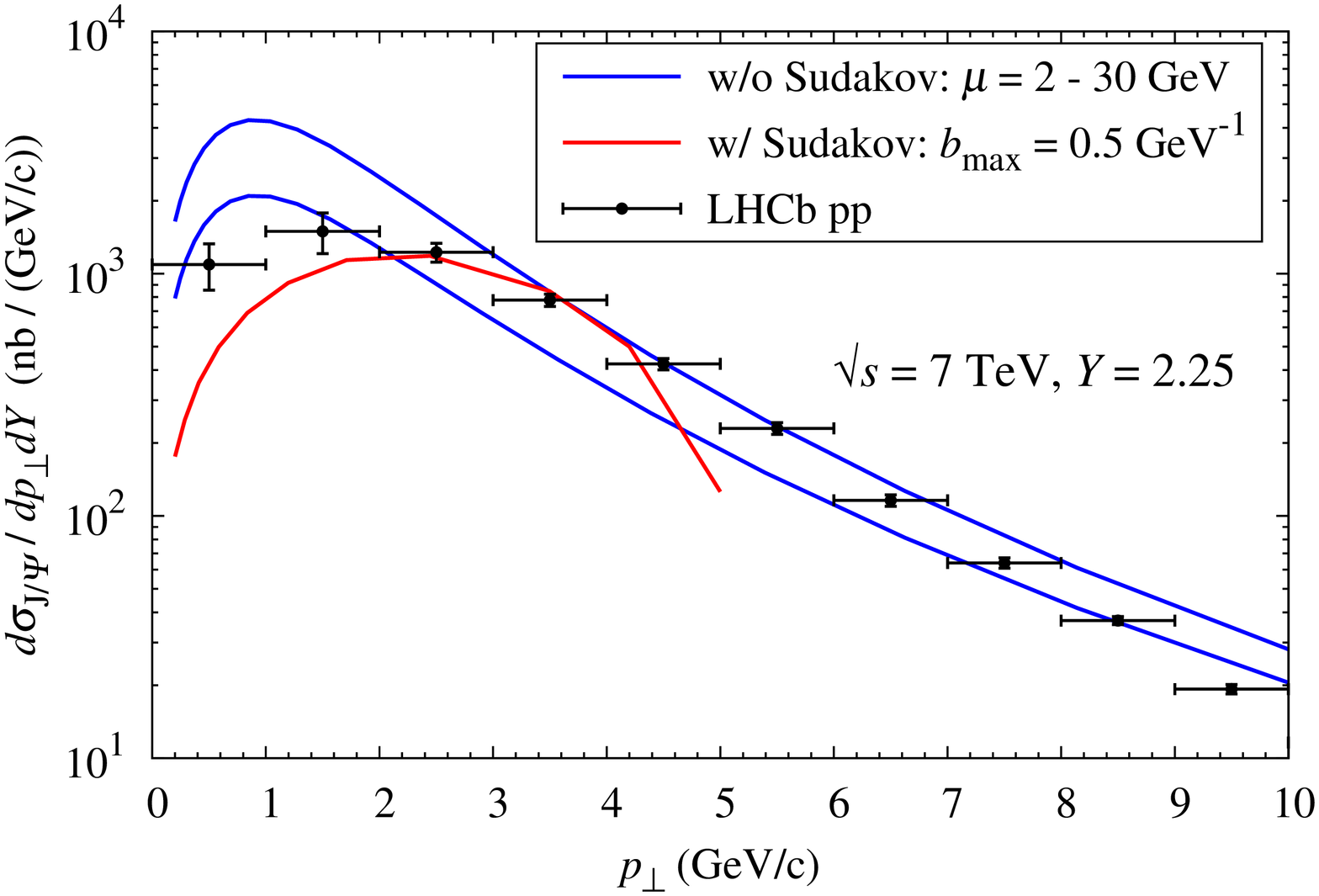} 
 \includegraphics[height=5.9cm,angle=270]{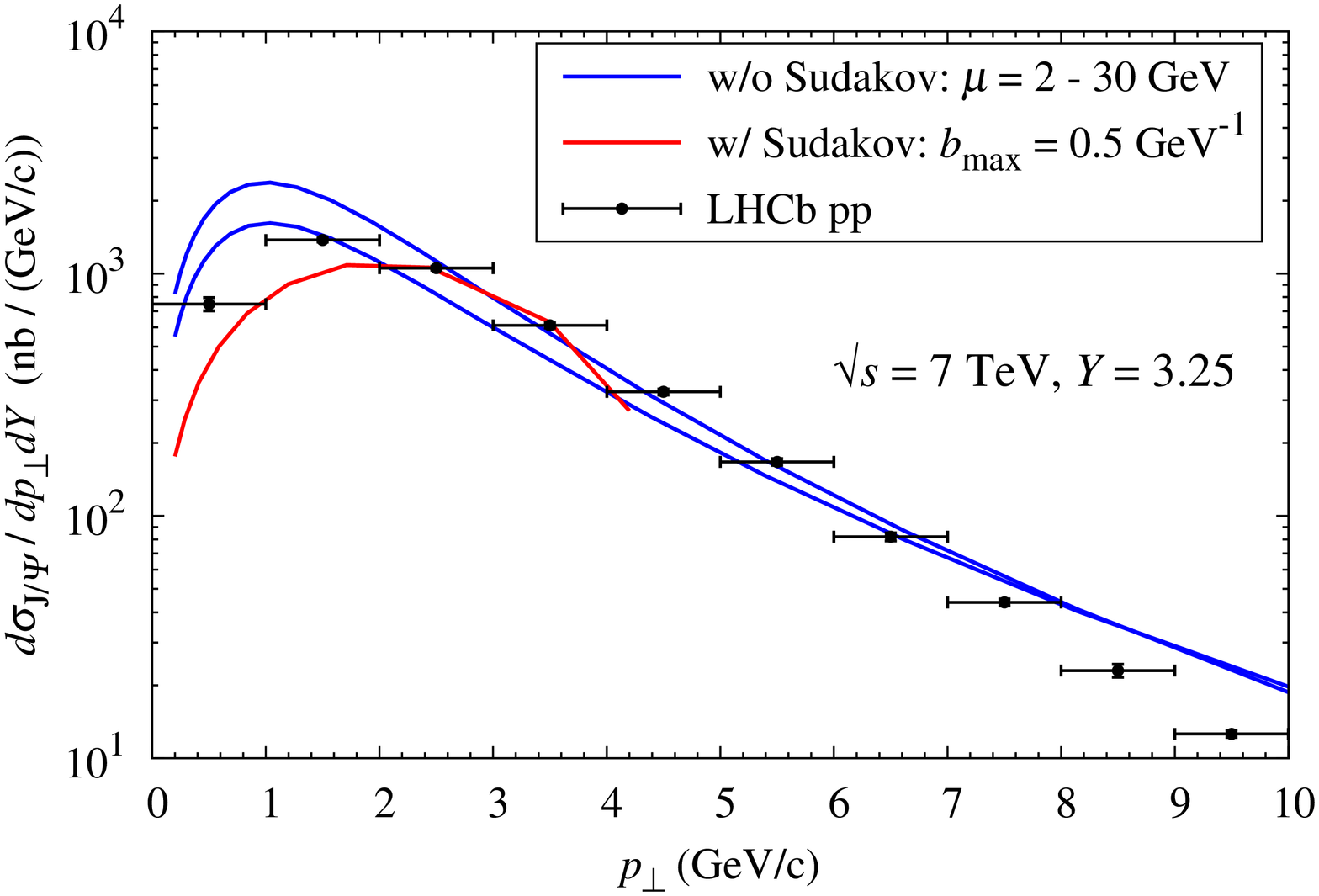}
 \includegraphics[height=5.9cm,angle=270]{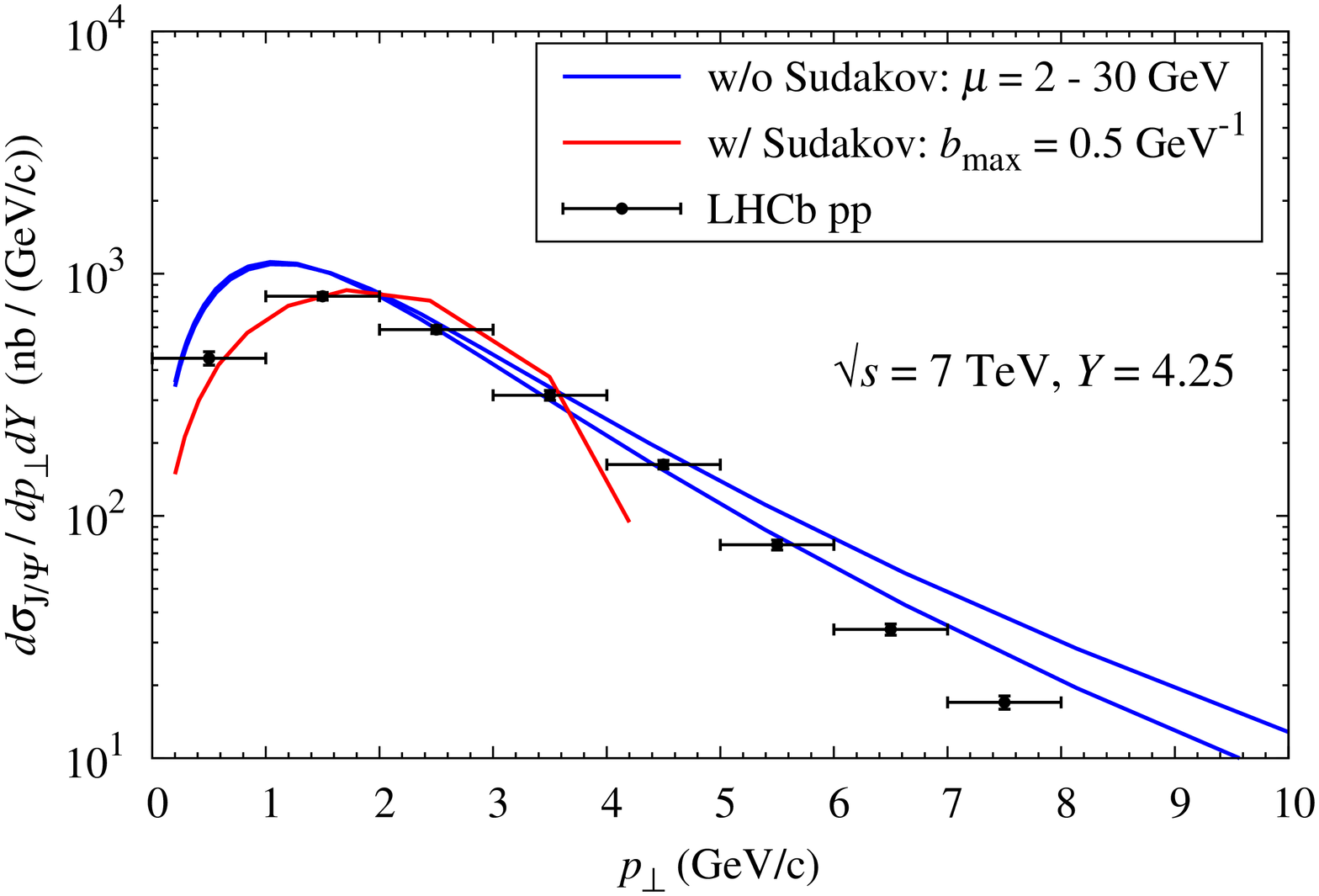} 
 \caption[*]{Double differential cross section of $J/\psi$ as a function of 
 $p_\perp$ for $Y=2.25$, 3.25, and 4.25 in pp collisions at $\sqrt s=7$ TeV. Blue solid line is obtained by using Eq.~(\ref{eq:cross-section-LN-coll}) and the uncertainty band is coming from a change of factorization scales ($2<\mu<30$ GeV) for the collinear gluon distribution function. Red solid line denotes the result of Eq.~\ref{eq:xsection_full} at $b_{\rm max}=0.5$.
We choose $F_{J/\psi}=0.0975$ for Eq.~(\ref{eq:cross-section-LN-coll}) and $0.1495$ for Eq.~(\ref{eq:xsection_full}). 
The LHCb data for prompt production is taken from Ref.~\cite{Aaij:2011jh}.
 }
 \label{results-jpsi}
\end{figure}

\begin{figure}[tbp]
 \centering
 \includegraphics[height=5.9cm,angle=270]{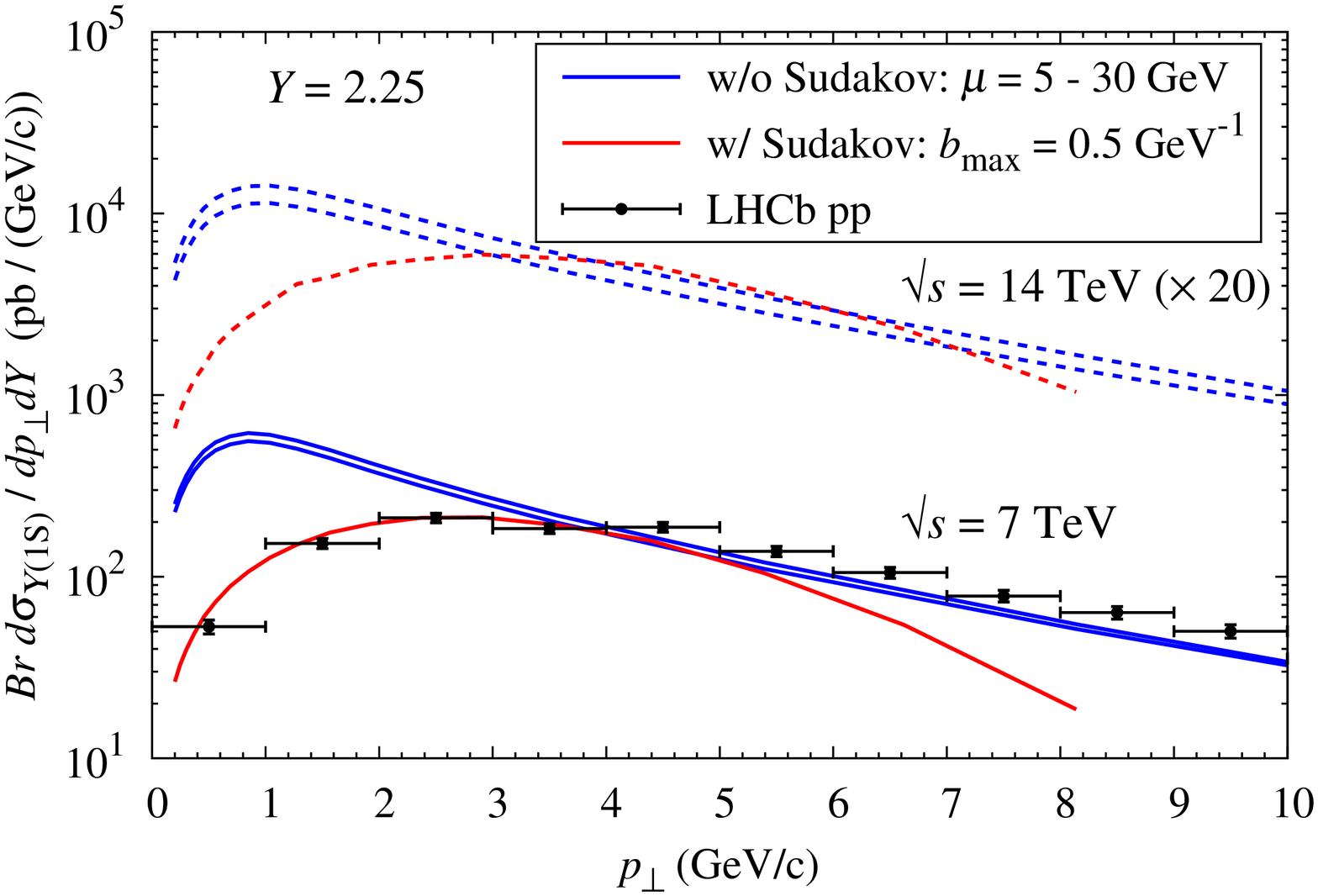} 
 \includegraphics[height=5.9cm,angle=270]{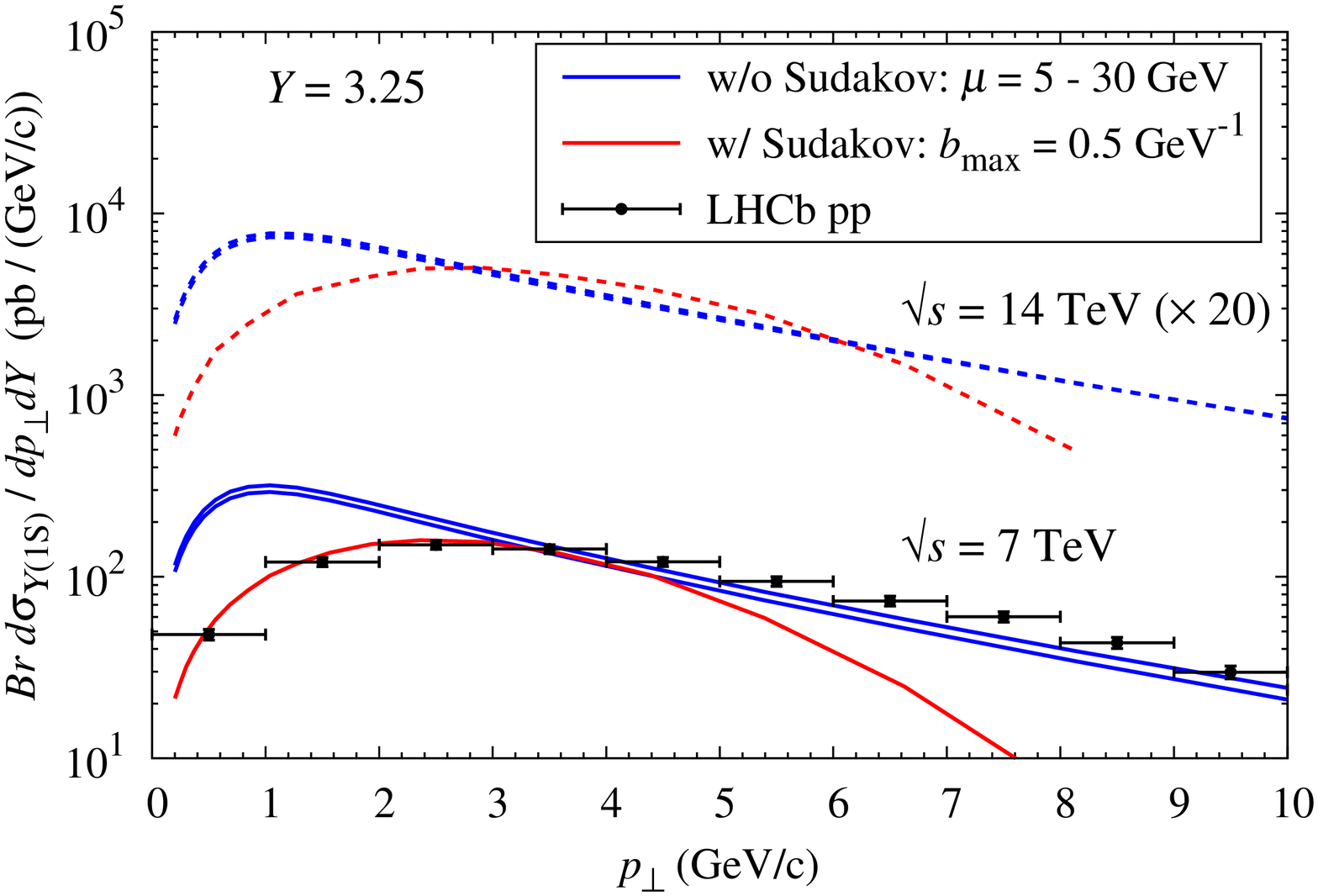} 
 \includegraphics[height=5.9cm,angle=270]{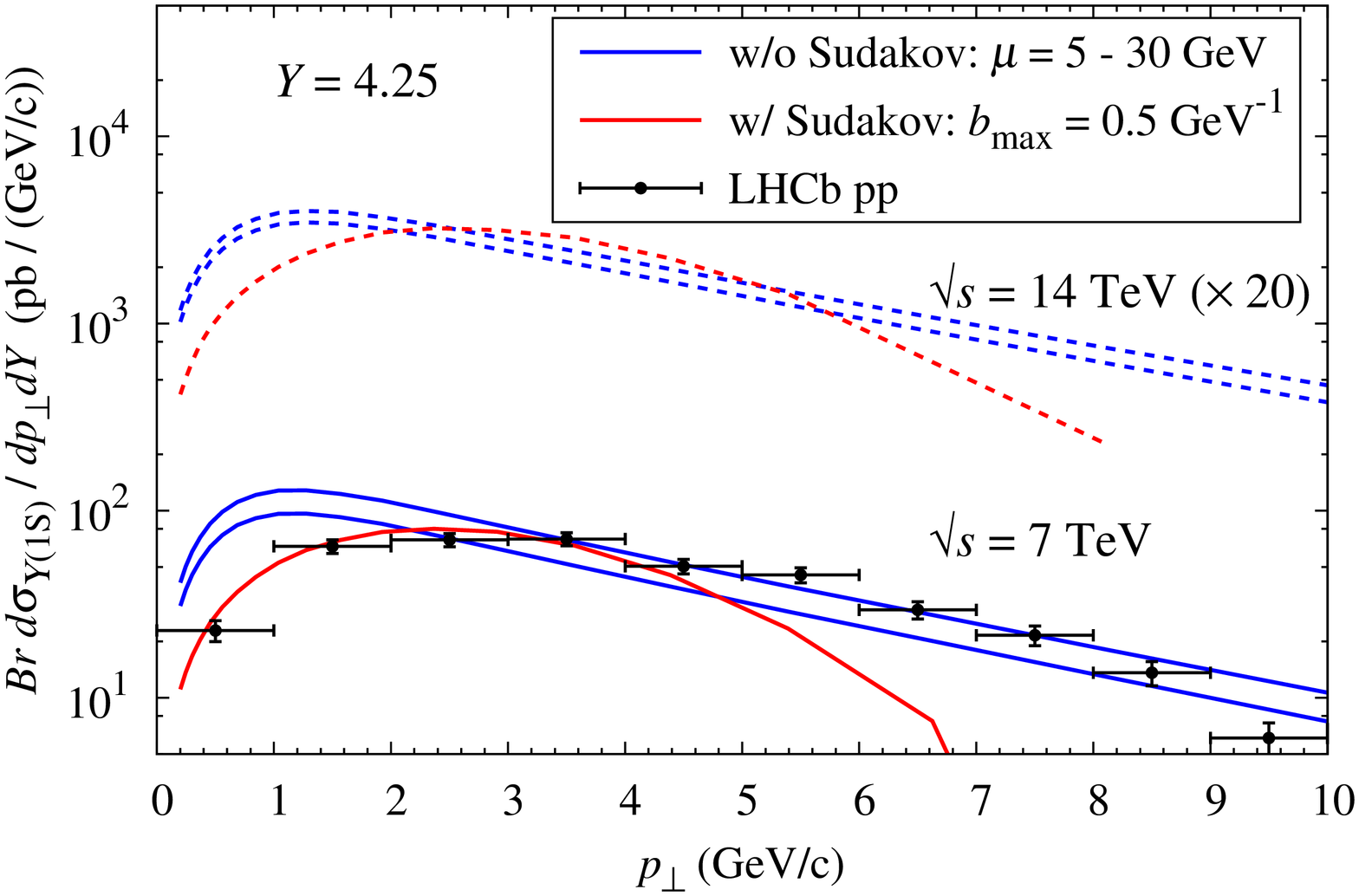} 
 \caption[*]{Double differential cross section of $\Upsilon$(1S) multiplied by a branching ratio of $\Upsilon$(1S) decay into a lepton pair as a function of $p_\perp$ for $Y=2.25$, 3.25, and 4.25 in pp collisions at $\sqrt s=7$ TeV (solid lines) and 14 TeV (dotted lines). 
We choose $F_{\Upsilon(1S)}=0.488$ for Eq.~(\ref{eq:cross-section-LN-coll}) and $0.390$ for Eq.~(\ref{eq:xsection_full}).
The LHCb data 
is taken from Ref.~\cite{LHCb:2012aa}. }
 \label{results-upsilon}
\end{figure}

\begin{figure}[tbp]
 \centering
 \includegraphics[height=5.9cm,angle=270]{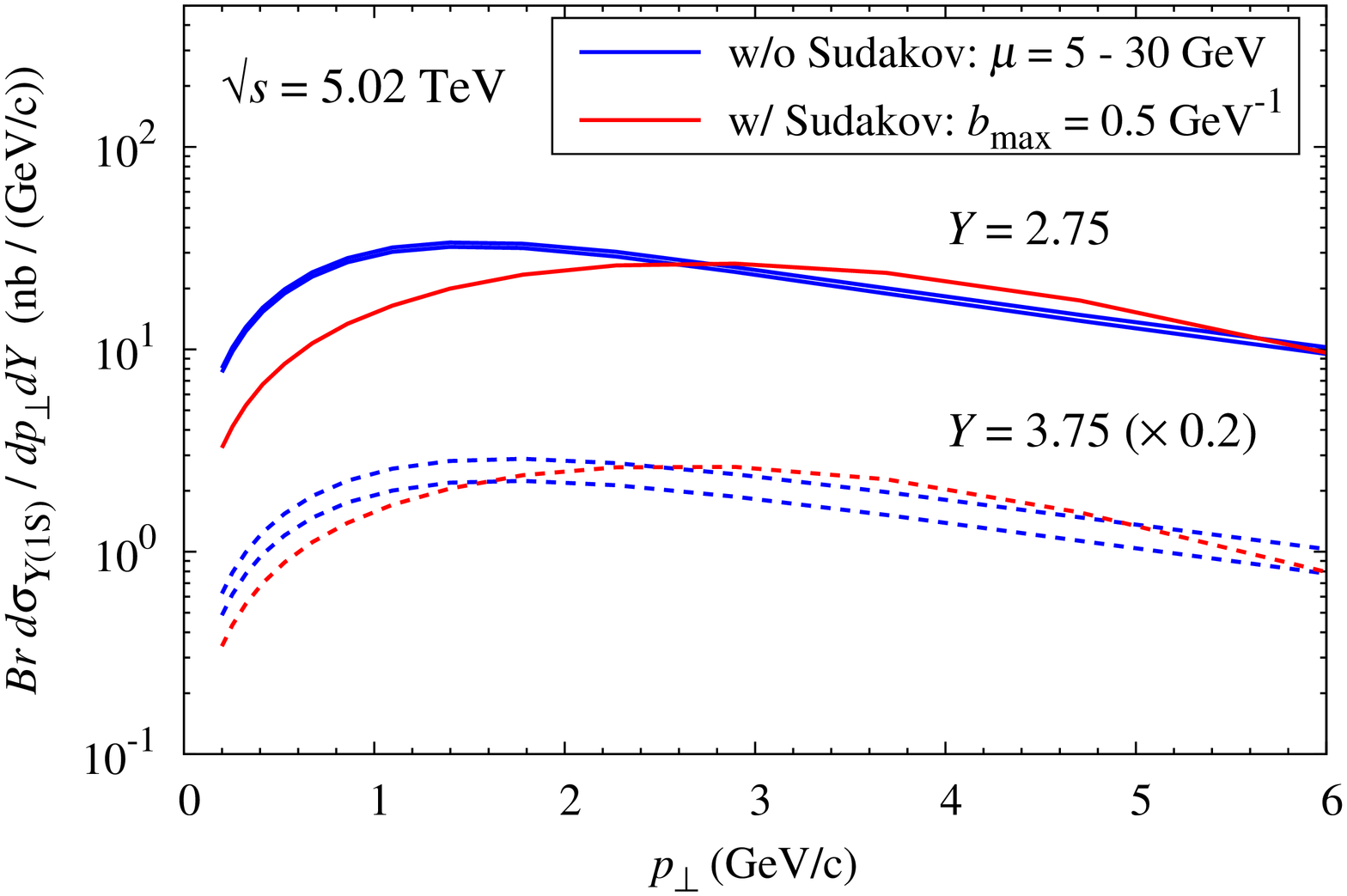} 
 \includegraphics[height=5.9cm,angle=270]{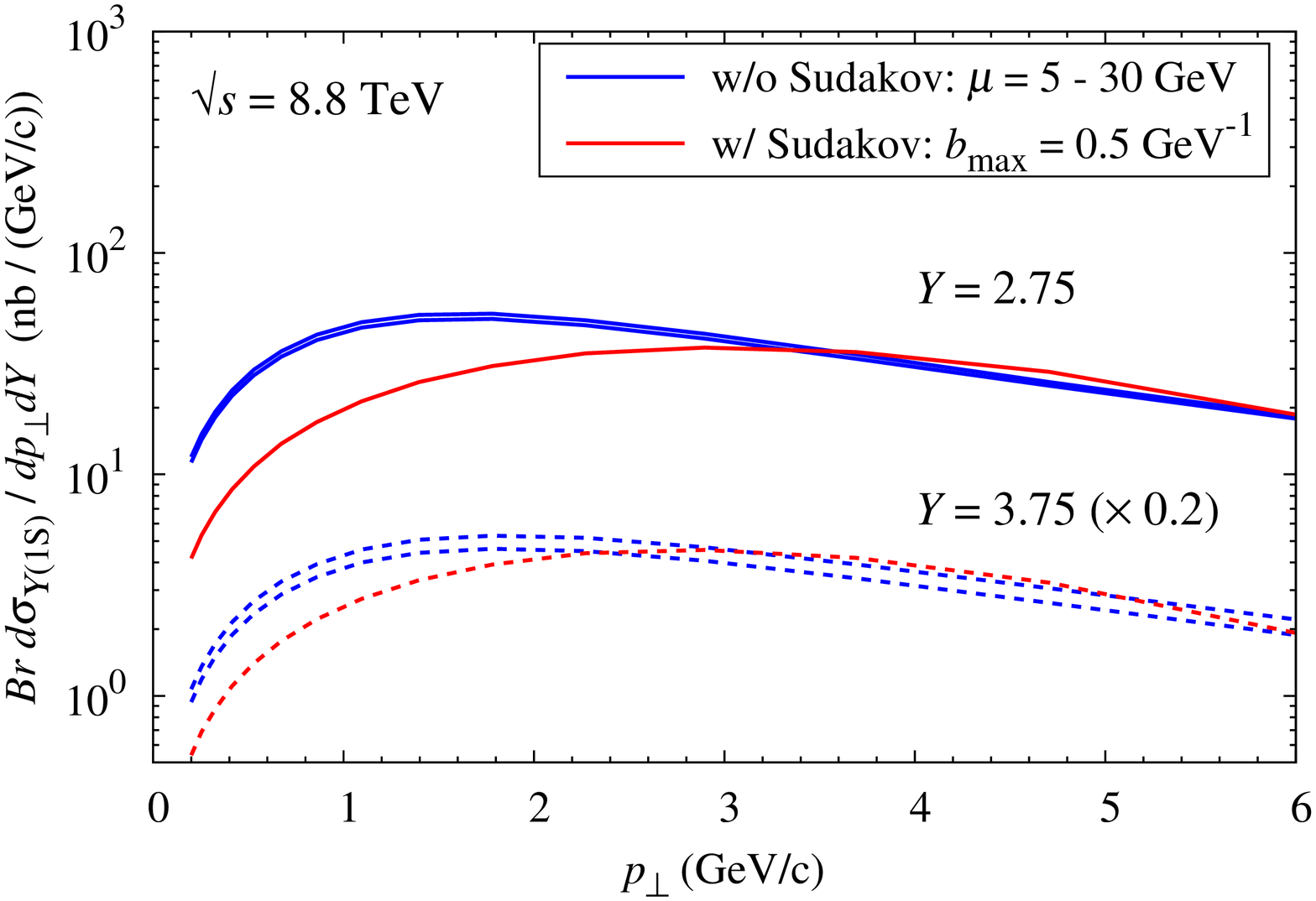} 
 \includegraphics[height=5.9cm,angle=270]{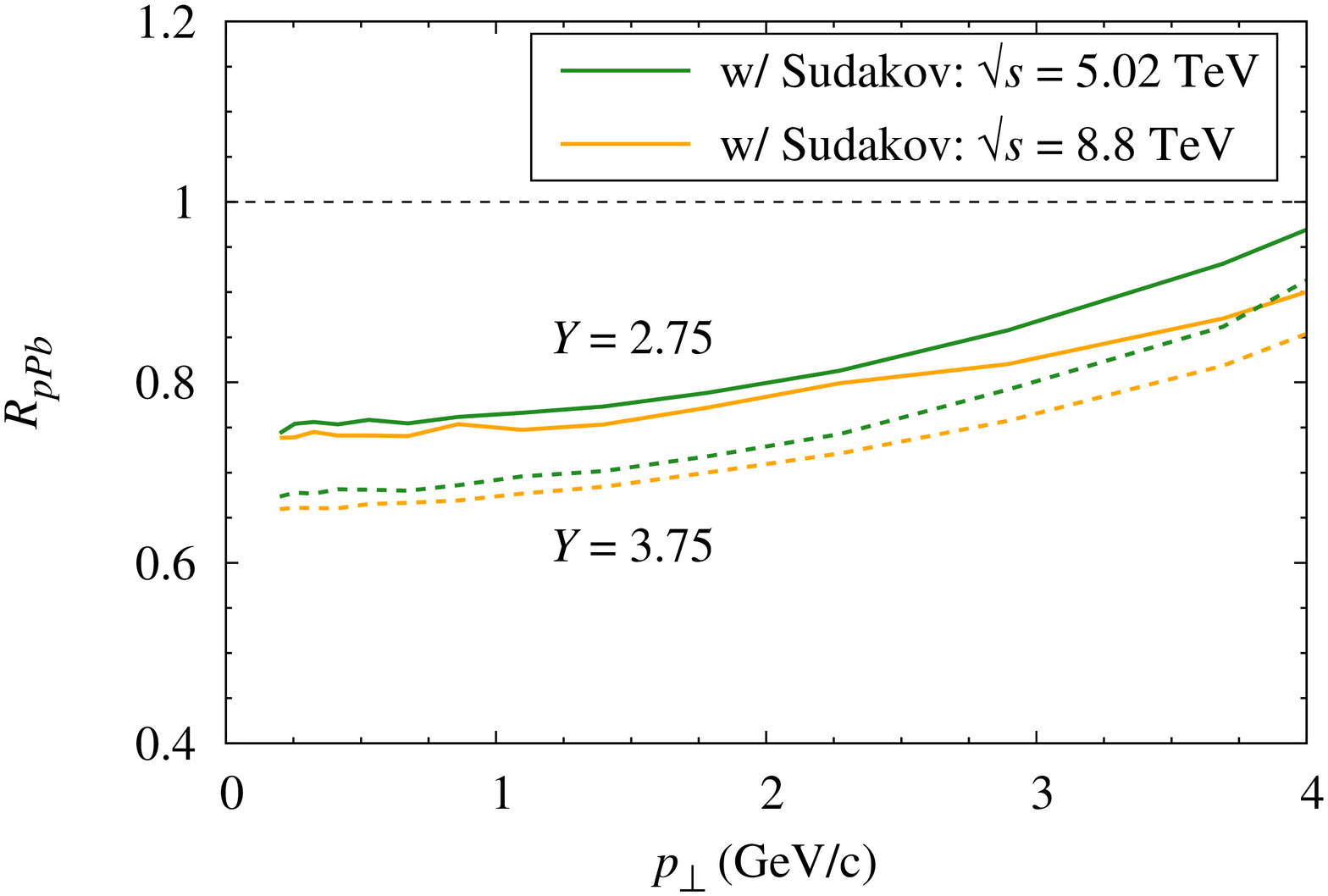}
 \caption[*]{Double differential cross section of $\Upsilon(1S)$  in $p{\rm Pb}$ collisions as a function of $p_\perp$ for $Y=2.75$ (solid lines) and $3.75$ (dotted lines) at $\sqrt s=5.02$ TeV (Left) and $\sqrt s=8.8$ TeV (Middle). 
Right: Nuclear modification factor of $\Upsilon(1S)$ as a function of $p_\perp$ at $5.02$ TeV and $8.8$ TeV for $Y=2.75$ (solid lines) and $3.75$ (dotted lines). 
 }
 \label{results-upsilon-pPb}
\end{figure}

The main focus of this paper is on the transverse momentum broadening of $\Upsilon$ production, while the parallel calculation for $J/\psi$ production is mainly treated as a comparison to illustrate the Sudakov effect.
We show in FIG.~\ref{results-jpsi} the differential cross section of prompt $J/\psi$ production as a function of 
$p_\perp$ in pp collisions at $\sqrt s=7$ TeV and at forward rapidity.
The results of Eq.~(\ref{eq:cross-section-LN-coll}), which is without the 
Sudakov effect, are computed with only small-$x$ evolution by using 
rcBK evolution. These results are roughly in agreement with the data, especially for $Y=4.25$, 
where the hybrid formalism is supposed to work the best. With the Sudakov factor, we find that Eq.~(\ref{eq:xsection_full}) provides slightly better 
agreement as compared with the result of Eq.~(\ref{eq:cross-section-LN-coll}). In general, 
the additional $p_\perp$ broadening provided by the Sudakov factor is negligible especially 
at $Y=4.25$ where the saturation effect is the strongest. 
We would like to emphasize that the Sudakov resummation is not applicable when $p_\perp$ is 
larger than the hard scale of the system. When $p_\perp \sim M$, we should 
switch to the fixed order CGC calculation, which is responsible for the large $p_\perp$ region of the spectrum.

FIG.~\ref{results-upsilon} shows the results for $\Upsilon$ production at $\sqrt s=7$ TeV and $14$ TeV. As shown in Refs.~\cite{Fujii:2013gxa,Ducloue:2015gfa}, the $\Upsilon$ production with only the small-$x$ 
resummation does not yield enough transverse momentum broadening as compared to data. We can find that the results from 
Eq.~(\ref{eq:xsection_full}) reproduce the data points very well due to the 
gluon cascade characterized by the Sudakov factor. 
The peak of the spectrum from Eq.~(\ref{eq:cross-section-LN-coll}) is located around $p_\perp=1$ GeV 
while the Sudakov factor shifts the peak position to around $3$ GeV.
Of particular importance is that the Sudakov factor in association with large $M$ gives 
additional strong broadening of the $p_\perp$ distributions for $\Upsilon$ production. 
The Sudakov resummation is indispensable in order to consistently describe all heavy 
quarkonium productions within the CGC/saturation formalism.

In FIG.~\ref{results-upsilon-pPb}, we finally show the differential cross section of $\Upsilon$ production as a function of $p_\perp$ in $p{\rm Pb}$ collisions at $\sqrt s=5.02$ TeV and $\sqrt s=8.8$ TeV by setting $Q_{sA,0}^2=3Q_{s,0}^2$ in the initial condition. We find that the Sudakov effect in $pA$ collisions is less pronounced as compared to $pp$ collisions, since the saturation effects becomes much stronger in $pA$ collisions due to the nuclear enhancement and it starts to dominate over the Sudakov effect. We also show in FIG.~\ref{results-upsilon-pPb} the $\Upsilon$ nuclear modification factor $R_{p{\rm Pb}}\equiv \frac{\sigma_{p{\rm Pb}}}{A \sigma_{pp}}$ as a function of $p_\perp$. The deviation of $R_{p{\rm Pb}}$ from unity can reveal the presence and strength of the gluon saturation effect. As shown in the $R_{p{\rm Pb}}$ plot, the suppression gets stronger in higher energy and more forward rapidity due to increased saturation effect.



\textit{Conclusion}--
In this paper, by computing the $p_\perp$ spectra for forward $J/\Psi$ and $\Upsilon$ productions at the LHC, we demonstrate that the Sudakov resummation on top of the small-$x$ evolution in the saturation formalism is essential to interpret the LHC data. Further theoretical and experimental efforts along this line can help us quantitatively study the $k_\perp$ dependent gluon distributions inside high energy protons and nuclei, and provide us insightful information about the hadron tomography especially in the low-$x$ limit, as well as important evidences for the onset of gluon saturation~\cite{Accardi:2012qut}. 

\textit{Acknowledgements}--BX would like to thank Dr. F. Yuan for discussions and the nuclear theory group at the Lawrence Berkeley National Laboratory for hospitality and support when this work is finalised.

\end{document}